\documentclass[twocolumn,aps,floats,floatfix]{revtex4}
\usepackage{graphicx,amssymb,color}
\usepackage[normalem]{ulem}
\usepackage{comment}

\begin{document}
\newcommand{\bb}{{\bf b}}
\newcommand{\beq}{\begin{equation}}
\newcommand{\eeq}{\end{equation}}
\newcommand{\bB}{{\mathbf{B}}}
\newcommand{\bE}{{\bf E}}
\newcommand{\bH}{{\bf H}}
\newcommand{\bD}{{\bf D}}
\newcommand{\bM}{{\bf M}}
\newcommand{\bN}{{\bf N}}
\newcommand{\bL}{{\bf L}}
\newcommand{\br}{{\bf r}}
\newcommand{\bV}{{\bf V}}
\newcommand{\murb}{{\bar{\mu}_r}}
\newcommand{\etheta}{{\bf e}_{\theta}}
\newcommand{\ephi}{{\bf e}_{\phi}}
\newcommand{\er}{{\bf e}_{r}}
\newcommand{\ex}{{\bf e}_{x}}
\newcommand{\ey}{{\bf e}_{y}}
\newcommand{\ez}{{\bf e}_{z}}
\newcommand{\ep}{{\bf e}_{+}}
\newcommand{\emm}{{\bf e}_{-}}
\newcommand{\eo}{{\bf e}_{0}}
\newcommand{\uvmn}{\,^{uv}_{mn}}
\newcommand{\mnuv}{\,^{mn}_{uv}}
\newcommand{\mnpq}{\,^{mn}_{pq}}
\newcommand{\dsum}{\displaystyle\sum}
\newcommand{\wtc}{\widetilde{c}}
\newcommand{\wtd}{\widetilde{d}}
\newcommand{\barc}{\bar{c}}
\newcommand{\bard}{\bar{d}}
\newcommand{\bark}{\bar{k}}
\newcommand{\tg}{\widetilde{g}}
\newcommand{\te}{\widetilde{e}}
\newcommand{\tf}{\widetilde{f}}
\newcommand{\barg}{\bar{g}}
\newcommand{\bare}{\bar{e}}
\newcommand{\barE}{\bar{E}}
\newcommand{\barf}{\bar{f}}
\newcommand{\breg}{\breve{g}}
\newcommand{\bree}{\breve{e}}
\newcommand{\bref}{\breve{f}}
\newcommand{\cG}{{\cal G}}
\newcommand{\cE}{{\cal E}}
\newcommand{\cF}{{\cal F}}

\newcommand{\tcb}[1]{\textcolor{black}{#1}}
\title{Nonlocal optical generation of spin and charge currents on the surface of magnetic insulators using total absorption and surface plasmons}
\author{S. T. Chui$^{1,2}$, Z. F. Lin$^{1,3}$, C.  R. Chang$^4$,
John Xiao$^2$}
\affiliation{
$^1$Bartol Research Institute, University of Delaware, Newark, DE 19716 \\
$^2$ Deptartment of Physics and Astronomy, University of Delaware, Newark, DE 19716 \\
$^3$ \tcb{Key Laboratory of Micro and Nano Photonic Structures (Ministry of Education),
     Fudan University, Shanghai, China}  \\
$^4$ Deptartment of Physics, Taiwan National University, Taipei, Taiwan }

\begin{abstract}
We study the nonlocal spin and charge current generation in a
finite metallic element
on the surface of magnetic insulators such as \tcb{yttrium iron garnet }
due to the absorption of the magnetic surface plasmon (MSP).
Whereas
a surface plasmon is completely reflected by a metal,
\tcb{an } MSP \tcb{can be } absorbed \tcb{due to the absence of backward states}.
The \tcb{injection of} MSP generates a voltage in the longitudinal direction
parallel to the wave vector, \tcb{with the voltage} proportional to
input power. If the metal is a ferromagnet, a spin current can also be
\tcb{induced }
in the longitudinal direction.
Our  \tcb{ results provide a way to improve upon
} integrated circuits of spintronics and spin wave logic devices.
\end{abstract}
\maketitle

PACS: 78.20.Ls,42.70.Qs,72.15.Gd,73.43.-f

Spintronics, which provides a promising way to overcome
the limitation of charged-based electronics
in data storage and processing, has received enormous attention recently \cite{spintronics}.
A crucial process in spintronics is the injection of spin and charge
currents between structures, at least one of which is magnetic. 
The injection current is reduced when there is an
impedance mismatch across the junction.
Improving the injection efficiency is of the utmost importance.
The injection can be local as in vertical structures
or nonlocal as in planar structures.
Nonlocal structures hold the promise
for large scale integrated circuits applications.
Current nonlocal spin transfer
structures \cite{nls} exploit diffusive spin currents.
In this structure, the distance 
travelled 
by the signal is equal
to the spin diffusion length $l_{sf}$,
and is
of the order of 1000 \AA.
Recently the focus of spin injection in vertical structures has shifted to
structures involving magnetic {\bf insulators}
such as 
yttrium iron garnet (YIG) \cite{fmr_yig},
where the loss may be smaller.

The exploitation of
magnetic surface plasmons (MSPs) in spintronics
have not been discussed previously but the MSPs offer
many intriguing potentials.
MSPs on bulk insulating magnetic surfaces
are also known as Damon-Esbach \cite{DE} magnetostatic modes. They are coupled modes of the {\bf surface}
spin wave and the electromagnetic (EM) wave and are the magnetic analog
of the better known (electric) surface plasmons. These states differ from bulk spin waves in that the frequency of these states are in the band gap region and corresponds to nonpropagating {\bf bulk} spin waves. They differ from
surface plasmons in that they are nonreciprocal, because of the
broken time reversal symmetry of magnetic systems.
This leads  to the absence of backscattering,
which is also exhibited by electronic "edge" states
in the
quantized Hall effect \cite{Bert} and in topological
insulators \cite{KM}. 
Stimulated by recent focus on plasmonics, there has
been interest in MSPs which can be generated
on the surface of magnetic insulators such as YIG \cite{yyr}.

When an EM  wave falls on a metal, it is
strongly reflected due to the high impedance mismatch;
very little signature of this wave is left in the metal.
A surface plasmon has an EM wave component and thus is completely
reflected as well when it strikes 
a metal. In plasmonics, the injection of surface plasmons into
metals has never been  considered.
Whereas a surface plasmon is nearly totally  reflected,
a MSP incident on a metal is absorbed
now that the reflected surface wave is inhibited by the broken time reversal symmetry.

In this Letter,
stimulated by a desire to improve upon the spin injection efficiency,
to reduce the loss of the signal over distance
of nonlocal structures, 
by
recent interests in spin wave logic devices \cite{swd}
and 
by experimental results of voltage generation in
a metal film M on  YIG under microwave
excitation \cite{expt},
we study the injection of an MSP into M (which may or may not be ferromagnetic),
as is illustrated in Fig. 1. We found that the
MSP can be  efficiently injected into M by the incoming electric 
and magnetic field regardless  of the interface ``impedance mismatch",
generating a static voltage due to the Hall effect in the metal and, also, a spin current if M is a ferromagnet.
This provides for the electrical detection of the MSP.
In addition, since the transport of the MSP is ballistic,
the loss of the signal over propagating distance is much reduced.
The distance travelled by the MSP is
{\bf macroscopic}, of the order of a centimeter. It is larger than the spin diffusion length
of current nonlocal spin valve structures by five orders of magnitude. Finally,
due to the partial spin wave nature of the excitation,
the MSP can be easily coupled to spin wave logic devices and provides an efficient connection between different circuit elements.
We now describe our results in detail.

\begin{figure}
\includegraphics[width=0.48\textwidth]{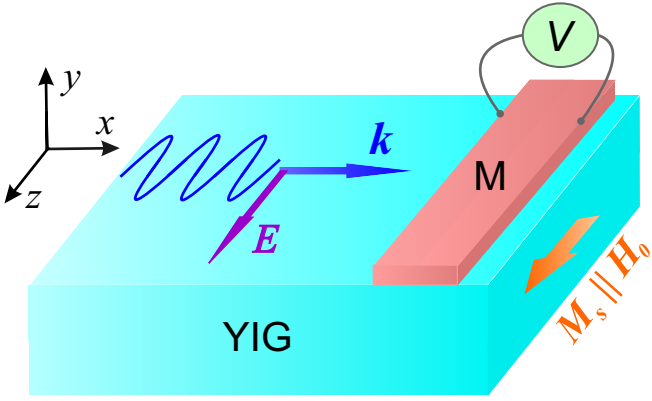}
\caption{\label{fig1}
(color online) Schematic diagram of the geometry for  the injection
of \tcb{an} MSP \tcb{into a metallic element M. The MSP has wave vector ${\bf k}$  along $x$
and the electric field E parallel
to the magnetization ${\bf M}_s.$ }}
\end{figure}

The nonlocal geometry we have in mind is illustrated in \tcb{Fig. 1}.
Magnetic surface plasmons can be generated at the surface of a magnetic
insulator with an external oscillating RF field, such as from a strip line, away
from a finite metallic element \tcb{M} deposited on the surface of a magnetic insulator. 
In the following we shall illustrate our
ideas with parameters such that the magnetic insulator is YIG.

\tcb{When fully magnetized, a} magnetic material is described by a magnetic
\tcb{permeability} tensor
\begin{equation}
\widehat{\mu}=\mu_0\left[%
\begin{array}{ccc}
\mu & -i\mu^{\prime} & 0 \\
i\mu^{\prime} & \mu & 0 \\
0 & 0 & 1 \\
\end{array}
\right],
\end{equation}
where $\mu$ \tcb{and} $\mu^{\prime }$ are of the  resonance form:
$\mu=1+f_m(f_0+i\alpha f)/[(f_0+i\alpha f)^2- f^2]$,
$\mu^{\prime}=\tcb{-} f_m f/[( f_0+i\alpha f)^2- f^2]$ with a
charcteristic bare
spin wave frequency $ f_0=\gamma H_0$. \tcb{Here} $H_0$ is a sum of the
external field $H_{ext}$ and the anisotropy field $H_a$, $\gamma$
is the
gyromagnetic ratio. $f_m=\gamma \tcb{\,4\pi M_s}$ 
measures the coupling strength of
the magnetic material with the EM waves and \tcb{$\alpha$ denotes the magnetic damping}.
For our calculation,
we use $H_0=900\,\text{Oe}$, $4\pi M_s=1750\,\text{Oe}$, \tcb{$\alpha=7\times10^{-3}$}, and the
permittivity 
$\epsilon_s=15+i7\times10^{-3}$.

We assume a geometry where the \tcb{YIG surface and the} film is in the $x-z$ plane with
\tcb{the applied magnetic field and} the magnetization $M_s$ along the $z$ direction.
We assume M to have
straight edges parallel to the $x$ and the $z$ axes.
The frequency of the
MSP at the air-YIG interface as a function of $k_x$ for $k_z=0$
for the above parameters is illustrated in Fig. 2.

There are four frequencies that are of interest. The bulk spin wave gap at long wavelengths occurs between the 
spin wave frequency $ f_{sw}=\sqrt{ f_0( f_0+ f_m)}$
and the frequency $f_b=f_0+f_m$. 
The MSP frequency
is in this band gap region where the bulk spin wave states are non-propagating. The other two frequencies are
$f_\nu=\sqrt{[\epsilon_s f_0(f_0+f_m)-f_0^2]/(\epsilon_s-1)}$,
where the dispersion with negative $k_x$ starts and
the frequency $f_{s}= f_0+ f_m/2$, 
at which the effective magnetic
permeability $\mu_{\text{eff}} =\mu-\mu'=-1$ and the dispersion
with positive $k_x$ ends.
For the above choice of parameters,
\tcb{they are given by
$f_{sw}=4.324$ GHz, $f_n=4.425$ GHz, $f_s=4.97$ GHz and $f_b=7.42$ GHz.}
The MSP frequency starts at $ f_{sw}$.
In the frequency ranges $ f_{sw}< f < \tcb{f_\nu}$ (yellow region)
and \tcb{$ f_s<f<f_b$} (blue region)
the frequency of the MSP is defined only for one sign of
the wave vector $k_x$ \cite{Hart}.
\begin{figure}
\includegraphics[width=0.48\textwidth]{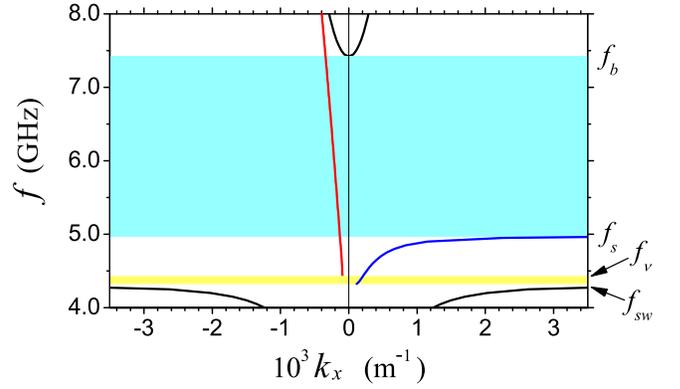}
\caption{\label{fig2}
(color online) 
The dispersion of the MSP at the surface of YIG. 
The yellow (cyan) region denotes the frequency range
where only MSP states with positive (negative) $k_x$ is allowed, 
while the MSP states with negative (positive) $k_x$ are inhibited, 
due to the the broken time reversal symmetry.
}
\end{figure}
Because of the broken time reversal
symmetry of magnetic systems at these frequency ranges there are no MSP states for wave
vectors of the other sign.
As can be seen from this figure, the group velocities $v_{MSP}$ of the two branches of the MSP are of the order of $10^8 cm/s$ and
the speed of light, respectively.
The distance travelled is controlled by the damping coefficient $\alpha$ and is
of the order $v_{MSP}/(\alpha \omega).$ For the two branches,
they range from 10cm to 1000 cm.
The \tcb{$y$} extent of the MSP,  $( (\omega)^2/c^2-k_x^2)^{-1/2},$
is of the order of a \tcb{millimeter}.

We next discuss the interaction of the MSP with the metallic film.
The usual formulation of spin and charge injection focuses on the matching of the spin and charge current at an interface. In the present case there are no charge current; the boundary occurs in air and there are {\bf no} discontinuity in the magnetic properties in the 
YIG substrate.
A MSP is a mixture of a spin wave and an EM wave localised near the interface.
The MSP is completely described by its EM field, the information of
the spin wave is implicitly included in the magnetic susceptibility.
The interaction of the MSP and M can thus be formulated in terms of
its EM fields.

When an EM field strikes a metal, usually
it is strongly reflected. This is what happens with the electric
surface plasmon.  For the MSP, there are no states with negative $k_x$s
that it can be reflected to.
We have considered the possibility that the MSP is stopped at the inteface,
then goes along the edges of M and around it. This will involve states with finite $k_z$ and with $k_x=0$.
We find that states with $k_z\neq 0$ are not possible. This
can be seen from the following physical argument. For a nonzero  $k_z$,
the transverse electric (TE) and the transverse magnetic (TM) modes are coupled;
the EM wave in YIG  have a finite component of the electric field $E_y$ in the
direction perpendicular to the film.  A finite $E_y$ induces electric charges at the interface and
implies that the state
will also have an electric surface plasmon character.
This is not possible unless one of the dielectric constants is negative.

We have also considered the possibility of the wave going up at the air-M boundary. To this end we have studied states at the air-YIG interface localized near M with imaginary in plane (along x) wave vectors but real wave vectors in the direction perpendicular to the interface (along y). We found that there are no states going up from the air-YIG interface. The only states possible are those coming in from above.
Thus there is no surface states for the incoming MSP to escape to and this state can be injected into M. 
In the following to illustrate our result we consider the
case with the MSP wave vector perpendicular to the direction of the
magnetization ($k_z=0$). The electric field is then along $z$. The
tangential electric and magnetic fields $E_z,$ $H_y$ and the perpendicular
field $B_x$ inside M is the same as that of the
incoming MSP as there are no reflected wave. We denote quantities in M and YIG with superscripts M and Y; the quantities in air have no supersctripts.
The tangential electric field inside M, \tcb{$E^M_z$}, generates a current with current density
given by $J^M_z=\sigma \tcb{E^M_z}$ where
$\sigma$ is the conductivity.
Because of the Hall effect,  the magnetic field along \tcb{$y$ } and
this current along z generate an electric field along \tcb{$x$ }
of magnitude
\begin{equation}
E^H_x=-R_H^M B_y^M J_z^M,
\end{equation}
where $R_H^M\approx -1 /(ne)$
($n$ is the density of the electrons)
is the Hall coefficient of M, \tcb{and the superscript $H$ denotes the $E$ field due to Hall effect.}
As we emphasize at the beginning of this paper, the fields only depend on the intrinsic
properties of the metal and are not functions of the interface properties and any possible
"impedance mismatch". 
The product of $E_z^M$ and $B_y^M$
and hence \tcb{$E^H_x$} possess finite
time independent components.
In general M has no electrical leads attached to it and is in an open
circuit situation. Charges will be induced at the surface until
an electric field of magnitude \tcb{$\langle E_x^H \rangle$}
is developed and the current stops flowing. This corresponds to a
voltage
$U_x=\int_0^{L_x} dx \tcb{\langle E_x^H\rangle}$
where $L_x$ is the length of M in the
$x$ direction.
The total voltage is controlled by the
spatial dependence of the EM wave inside M. We show next that the electric field decays in M with a length scale of the order of an effective skin depth.

The electric field in a medium is described by Maxwell's equation
 $\Omega E(r) =0$
where
$ \Omega=\nabla \times \mu^{-1}/\epsilon \nabla \times -  \omega^2;$
$\mu$ is the magnetic \tcb{permeability }; $\epsilon$,
the dielectric constant. Here we have used units so that the speed of light $c$=1. In principle, the electric field inside M/YIG can be calculated from its value
at the boundary with a Green's function
technique \cite{gf}
from the equation
\begin{equation}
E(r)=-\int dS\cdot [ (\mu^{-1}/\epsilon \nabla \times E(r_S)) \times G(r,r_S)^*
\end{equation}
$$
- (\mu^{-1}/\epsilon \nabla \times G(r,r_S))^*\times E(r_S) ].
$$
where the surface integral is carried out at the YIG-air YIG-M boundary. $G$ is the Green's function defined by:
$
 \Omega' G(r,r')=\delta(r'-r).
$
The operator $ \Omega'= \Omega_{M}$ for $y>0,$
$ \Omega'= \Omega_{YIG}$ for $y<0.$ $G$ can be constructed from a linear combination of products of photonic states including the M-YIG MSP states 
as $G=\sum_{k_x}|n><n|/(k_0^2- \omega_n^2(k_x))$ where
$\omega_n$, $|n>$ are the eigenvalues and the eigenstates of $\Omega'$.
The Green's function is dominated
by contributions from the MSP states at the air-M-YIG  interface with energy $\omega_n(k_x)=k_0^2$
when the denominator is zero.
Thus the spatial dependence will be determined by these states which we study next.
To illustrate the
physics we first consider the simpler case where the thickness of
M is larger than the spatial extent of the MSP.
The MSP state  at the air-YIG interface is coupled to the
MSP state at the M-YIG interface which can be constructed as follows.

{\it MSP for M/YIG:} The properties of the interface MSP state is obtained by matching
$H_{\parallel}$ and $B_{\perp}$ of the plane wave states of YIG and
that in  M. 
We obtain the dispersion for the YIG-M MSP state given by 
\begin{equation}
k_{xM}^2 \approx - \omega^2\epsilon_{M}/[(b'-b+a)^2/a'-a'].
\label{disp1}
\end{equation}
Here $a=\mu_{x,x}^{-1}(YIG),$ $b=\mu_{x,y}^{-1}(YIG);$
$a'=\mu_{x,x}^{-1}(M)$ ,$b'=\mu_{x,y}^{-1}(M).$
$\epsilon$, $\epsilon_M$, are the dielectric constant of YIG and M.
These states can be excited by the incoming MSP and decays
exponentially with a length scale of the order of inverse skin depth $1/\kappa
=1/{\rm Im}[ (\omega\epsilon_M/a')^{1/2}]$
away from the air/YIG-M/YIG boundary. 


{\it MSP for a trilayer structure:}
So far we have assumed that the thickness of M is larger than
the spatial extent $d_y$ of the MSP. 
$d_y$ is of the order of a mm.
Current experiments are carried out with the thickness $t$ of the metallic film of the order
of nm and much less than $d_y$.
It is necessary to take the finite
thickness of M into account.

We have calculated the interface states in the air/M/YIG trilayer
structure by matching the boundary EM fields at both the air/M and
the M/YIG interfaces. We found that in the limit when
$t<<1/\kappa$, $k_x$ as
given by Eq. (\ref{disp1}) is modified so that $\epsilon_M$ is replaced by an effective dielectric constant that is
reduced by an amount of the order of $t\kappa.$
Furthermore, the real and the imaginary parts of $k_x$ are opposite
in sign. 
In place of \tcb{Eq. (\ref{disp1}) }  the dispersion is now given by

$$
k_{xM}/[\omega^2\epsilon_M t]\approx -1 /(1+a+b); 1/(1+a-b).
$$
The two solutions have opposite signs 
for the real part of $k_x.$ 
The voltage generated is determined by
the imaginary part of the wave vector and is given by
\beq
U_x= R_h\sigma B_yE_z min(l,L_x)
\eeq
where $l^{-1}\approx 2 \omega^2 Im[\epsilon_M] t
/(1+a-b)$ is of the order $t\kappa^2.$ For $t\approx 10nm$ and a skin depth $\kappa^{-1}$
of the order of a micron, we obtain $l\approx 0.1 mm.$ 
We next estimate the magnitude of this effect.

We take M to be Au with a Hall coefficient
$R_H=7.25 \times 10^{-11} m^3/C,$ and a conductivity $\sigma=4.55\times 10^7 /(ohm-m).$
Thus \tcb{$E_x^H =3.3\times 10^{-3} B_y E_z$ }.
We estimate the product $BE$ from the power $P$ of the input coplanar
waveguide.
We assume that most of the power is expanded in generating the MSP.
The energy current \beq
S=EB/\mu_0=P/A
\label{power}
\eeq
where $A$ is the area of the
wave.
For current experimental systems, the width in the z direction is of
the order of a mm; the y extent of the MSP is also of the order of a mm. Thus
$A\approx 1 mm^2.$ We obtain $<E_zB_y>\approx 4\pi P c/(10 v_{MSP}).$
Using the above estimate of $l$ and for $P$ of the order of a watt, $U_x$
is of the order of a micro-volt.


Most metals are paramagnets. Because of YIG, or in the presence of
an external magnetic field, the electrons in M has a small
magnetization. A nonferromagnetic metal usually do not possess local
moments and we do not expect exchange coupling between the YIG spins
and this small magnetization in the paramagnet.
The electric field $E_z$ generates an AC current inside M.
As the electric field oscillates from parallel to opposite the direction of
the induced magnetization,
a residual DC current can also generated if the
resistances along the two directions are different.
This can happen
if there is a significant amount of spin orbit impurity scattering.
The resistance along the two directions will then be different,
resulting in a net average current/voltage in this
transverse direction.
We next consider the situation when the metal is a ferromagnet.

The Hall coefficient $R_H$ of ferromagnets contains a component due to the anomalous Hall effect and the magnitude of the above effect can be modified. In ferromagnets a spin current is also induced.
At the air/YIG-M/YIG boundary, the perpendicular component of the magnetic
field $B_x$ and the tangential component 
$H_y$ are continuous.
Whereas the magnetization density in air is zero, there is a finite
change in the x, y components of the magnetization density at the
boundary inside the ferromagnet. We can obtain the magnetization
density from the equation
\tcb{$ {\bf M}(M) =[{\bf B}(M)/\mu_0- {\bf H}(M) ]$. }
$M_{x,y}$ act as sources to create  magnetization
currents in the ferromagnet. 
From these  we get the magnetization current given by $J_{\bf M}=i\omega {\bf M}$ with
\begin{equation}
M_x=
(a'-1/\mu_0-b'^2/a')B_x+ib'B_y/(\mu_0a'),  
\end{equation}
\begin{equation}
M_y=\left \{ B_y[1-1/(\mu_0a')]-ib'B_x/a' \right \} /\mu_0.
\end{equation}

Associated with the change in $M_{x,y}$ there is a static change in the longitudinal magnetization $M_z$
to second order: $\Delta M_z=\sqrt{M_0^2-M_x^2-M_y^2}-M_0$.
This induces an additional charge current given by\cite{fmrprb} 
$J'=-D_M\nabla(\Delta M_z)$ where $D_M$ is the diffusion constant  that relates the charge current to the longitudinal magnetization change.
In the open circuit situation, this provides for an additional contribution to the DC potential given by 
\beq
U\approx D_M <M_x^2+M_y^2>/(2M_0\sigma).
\eeq
We next estimate the magnitude of $U$ as a function of frequency.

\begin{figure}
\includegraphics[width=0.48\textwidth]{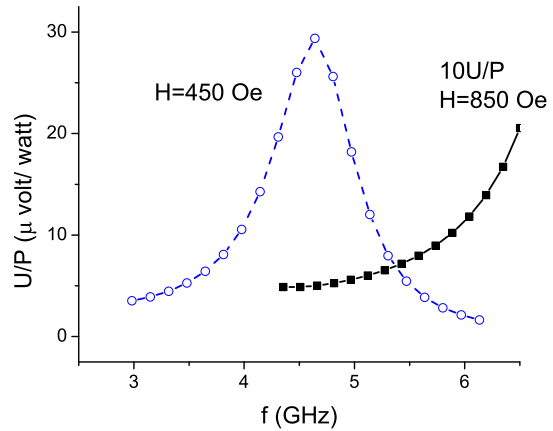}
\caption{\label{figu}
(color online) The output voltage from the spin current
normalized by the incoming power as a function of frequency for the higher frequency branch of the MSP for external magnetic fields $H_{ext}=450 Oe$ (dotted line) and $H_{ext}=850 Oe$ (solid line).}
\end{figure}

The ratio $D_M/\sigma$ is of the order of $1/(N(E_F)\mu_B e)$ where $N(E_F)$ is the density of states at the Fermi surface; $\mu_B$, the Bohr magneton; $e$, the electric charge.
From eq. (\ref{power}) we get $<B_y^2>=P\mu_0/(v_{MSP}A).$ 
We take M to be permalloy with a density of states $N(E_F)\approx 3\times 10^{28}/ev/m^3.$ For the spin wave parameters, we found recently\cite{Chen} that the fmr frequency of thin permalloy film as a function of frequency and external field $H_{ext}$ can be fitted by the typical spin wave formula $f_{sw}$ with satutaion magnetization $\mu_0M_0=1 T,$ and total field 
$H_0=H_{ext}-126.6 Oe.$ Our estimates for the voltage U normalized
by the incoming power P as a function of frequency for the upper branch of the MSP are shown
in fig.\ref{figu} for external fields
of magnitudes 850 Oe and 450 Oe. The results for $H_{ext}=850$ Oe are multiplied by a factor of 10. As can be seen, the magnitude of this contribution  is of the order of microvolts. It can exhibit a peak as a function of frequency. This peak is due to the fmr resonance of the permalloy metallic film when $a'$ is close to zero. The magnitude of this peak depends on the Gilbert damping parameter which we have set to a value of 0.03.

It is likely that the spins of YIG and that of the
ferromagnet are exchange coupled. The continuity of $H_x$
and $B_y$ across the YIG-ferromagnet interface imposes partial alignment
of the spins between these two materials. Spin currents in the FM will
generate some residual spin wave in YIG as well. In the present calculation the exchange coupling
between spins can be incorporated
in the magnetic susceptibility in eq. (1) by including 
the exchange energy in $ f_0$.

In conclusion we
study the nonlocal spin and charge current generation in a
finite metallic element
on the surface of magnetic insulators
with the magnetic surface plasmon (MSP),
which is absorbed and cannot be reflected.
The MSP generates a voltage in the longitudinal direction  proportional to
the power. If the metal is a ferromagnet, a spin current can also be generated
in the longitudinal direction.
Our result will be useful for new classes of devices using spin waves.

Acknowledgement: John Xiao is supported by DOE under grant number DE-FG02-07ER46374.
\tcb{ZFL is partly supported by the 973 program (2011CB922004).}

\end{document}